# Growth and hydrostatic-pressure study of a type-II superconductor $Bi_2Ta_3S_6$ single crystal

Chenglin Li[1, 2 =], Yaling Yang[1, 2 =], Zhilong Yang[3 =], Junze Deng[4], Ruihan Zhang[1,2], Weiwei Chen[1, 2], Yue Pan[1, 2], Yulong Wang[1, 2], Xuhui Wang[1, 2], Bosen Wang[1, 2], Zhijun Wang[1, 5]* and Gang Wang[1]*

[1]*Beijing National Laboratory for Condensed Matter Physics, Institute of Physics, Chinese Academy of Sciences, Beijing 100190, China*

[2]*School of Physical Sciences, University of Chinese Academy of Sciences, Beijing 100049, China*

[3]*School of Mathematics and Physics, University of Science and Technology Beijing, Beijing 100083, China*

[4]*Department of Applied Physics, Aalto University School of Science, Aalto FI-00076, Finland*

[5]*Condensed Matter Physics Data Center, Chinese Academy of Sciences, Beijing 100190, China*

= These authors contributed equally to this work.
*Corresponding authors: <u>wzj@iphy.ac.cn</u>; <u>gangwang@iphy.ac.cn</u>;

# Abstract

We report the growth and physical properties of single-crystalline $Bi_2Ta_3S_6$ crystallizing in $P6_3/mcm$ space group, which comprises alternating Ta-S layers and Bi layers with each Bi atom connected with adjacent S atoms. Temperature-dependent electrical resistivity measurements reveal a superconducting transition at 0.84 K, with upper critical field 231 Oe under an out-of-plane magnetic field. The magnetization measurements confirm its nature as a type-II superconductor, with anisotropic Ginzburg-Landau parameter $\kappa_{ab}$ = 7.67 and $\kappa_c$ = 4.50. Hall measurements indicate the dominant carriers as hole. Hydrostatic pressure is applied, under which both the superconducting transition temperature and upper critical field increase sharply under low pressure before undergoing slight suppression under higher pressure. Density functional theory calculations reveal non-trivial topological surface states on (100) surface in $Bi_2Ta_3S_6$, which may offer a new avenue for exploring potential topological superconductivity in layered transition metal dichalcogenides.

# Introduction

Topological superconductors (TSCs) have emerged as promising platforms for next generation fault-tolerant quantum computers due to their hosting of Majorana fermions which obey non-Abelian statistics [1]. Theoretical calculations proposed that chiral *p*-wave superconductors, such as $Sr_2RuO_4$, as TSCs candidates [2-6], though their topological nature requires further experimental verification. Alternative strategies in search of TSCs include: inducing superconductivity in topological insulators/semimetals via chemical doping or applying pressure [7,8]; constructing heterostructures combining topological insulating layers and superconducting layers [9,10]. However, these approaches face their own challenges such as achieving uniform doping, conducting high-pressure spectroscopic studies, or overcoming lattice mismatches in artificial heterostructures. Therefore, seeking intrinsic TSCs that inherently integrating superconductivity with topological band structures represents a more promising avenue for realizing robust topological superconductivity [11].

$ATaX_2$-type compounds (A = Pb, Sn, In, and Bi; X = S, Se) have garnered considerable attention due to their intrinsic nature of superconductivity and non-trivial topological band structures [12-16]. These compounds adopt layered structures hosting alternating $TaX_2$ layers and A atomic layers. Structurally, $ATaX_2$ compounds can be categorized into two groups, as one with non-centrosymmetric $P\bar{6}m2$ space group, $PbTaSe_2$ for example, and another with centrosymmetric $P6_3/mmc$ space group, such as $PbTaS_2$ [17]. Theoretical calculations reveal that the hybridization of Ta atoms d-orbitals and A atoms *p*-orbitals in $ATaX_2$ induces the formation of nontrivial topological bands [18-21]. Experimentally, intrinsic superconductivity has been observed in $ATaX_2$ compounds such as $InTaS_2$ [14,15,18], $PbTaSe_2$ [12,13,20-22], $PbTaS_2$ [17,23], and $SnTaS_2$ [16,24], which inspires us to search for topological superconductivity in $ATaX_2$-type materials. $PbTaSe_2$ is a topological nodal-line semimetal with a superconducting transition at 3.8 K [21,22]. Angle-resolved photoemission spectroscopy measurements on PbTaSe$_2$ reveal its nodal-line features in the energy bands near the Fermi level protected by mirror symmetry [21]. Meanwhile, linear magnetoresistance was observed in PbTaSe$_2$ under a magnetic field of 0 T - 2 T, potentially related to its non-trivial topological bands [22]. Notably, zero-bias conductance peaks were detected via temperature-dependent spectroscopic imaging-scanning tunneling microscopy measurements, suggesting it as a candidate for TSCs [12]. For PbTaS$_2$, bulk superconductivity was observed at 2.6 K. First-principles calculations indicate that the energy bands of PbTaS$_2$ possesses topological nodal line without spin-orbit coupling (SOC) [23]. However, when SOC is considered, a bandgap opens at the nodal-line states, leaving only topological surface states on specific surface.

In this work, we report the first growth of high-quality Bi$_2$Ta$_3$S$_6$ single crystals using a chemical vapor transportation (CVT) method. Bi$_2$Ta$_3$S$_6$ shows a two-dimensional (2D) characteristic, where Bi ions are intercalated in between Ta-S layers and site at the top of S atoms. Electrical transport and magnetization measurements reveal a superconducting transition with $T_c$ = 0.84 K under ambient pressure. Furthermore, magnetization measurements suggest Bi$_2$Ta$_3$S$_6$ as a type-II superconductor. Hydrostatic pressure was adopted as an effective method to tune the superconductivity, with $T_c$ and $H_{c2}$ enhanced to 4 times and 4.5 times compared to the ones at ambient pressure, respectively. First-principles calculations reveal that the states near the Fermi level in Bi$_2$T$_3$S$_6$ are primarily contributed by the Bi atoms and Ta atoms. Non-trivial topological Dirac points were predicted in bulk states without SOC. When SOC is considered, Bi$_2$T$_3$S$_6$ exhibits non-trivial surface states on (100) surface, providing a possibility for potential topological superconductivity.

## Experimental Methods

Bi$_2$Ta$_3$S$_6$ single crystals were synthesized via a CVT method using iodine as transport agent. High-purity Bi (powder, Alfa Aesar, 99.999%), Ta (powder, Innochem, 99.9%), and S (powder, Innochem, 99.95%) were weighted in a molar ratio of 2 : 3 : 6 and total mass of 2.0 g, and homogenized thorough grinding. The mixture was placed in a vacuumed quartz tube with 100 mg iodine, which was subsequently flamed-sealed and then transferred into a horizontal two-zone furnace with a thermal gradient. The source zone and crystal growth zone were maintained at 1323 K and 1173 K for a week, respectively. After natural cooling to room temperature, hexagonal plate-like crystals with metallic luster were obtained.

Powder X-ray diffraction (PXRD) data was acquired on a PANalytical X'Pert PRO diffractometer (Cu $K_{\alpha 1}$ radiation, $\lambda$ = 1.54056 Å) operated at 40 kV voltage and 40 mA current with a graphite monochromator in a reflection mode ($2\theta$ = 10 - 90°, step 0.017°). Single-crystal X-ray diffraction (SCXRD) patterns of Bi$_2$Ta$_3$S$_6$ were collected by a Bruker D8 VENTURE single crystal X-ray diffractometer with Mo $K_\alpha$ radiation ($\lambda$ = 0.71073 Å) at ambient temperature. The crystal structure was solved by a direct method and refined via a full-matrix method employing the SHELXTL software package in Olex2 software [25]. The semi-quantitative elemental analysis was carried out using an energy dispersive X-ray spectroscopy (EDS) on cleavage planes. The atomic-scale images were acquired using a JEM-ARM 200CF scanning transmission electron microscopy (STEM) operated at 200 kV with a STEM high-angle annular dark-field (HAADF) imaging. The specimens for STEM were prepared via careful grinding.

Temperature-dependent electrical transport measurements of $Bi_2Ta_3S_6$ were conducted in a physical property measurement system (PPMS-16T, Quantum Design), and field-dependent magnetization measurements were carried out via a vibrating sample magnetometer. Ultra-low-temperature transport measurements were conducted on a dilution refrigerator. The superconductivity was investigated on a magnetic property measurement system (MPMS-3 EverCool, Quantum Design). Temperature-dependent resistivity under hydrostatic pressure was measured using a self-clamped palm-type cubic anvil cell [26]. Liquid glycerol was used as the pressure transmitting medium (PTM). The standard four-probe method was employed with the current applied within the *ab* plane and the magnetic field parallel to the *c* axis. All hydrostatic-pressure measurements were performed in a $^4$He refrigerated cryostat with a 9 T superconducting magnet down to 1.6 K at the Synergic Extreme Condition User Facility, Huairou, Beijing.

First-principles calculations were performed based on density functional theory (DFT), using the projector augmented wave method [27,28] implemented in the Vienna *ab initio* simulation Package [29]. The generalized gradient approximation was adopted for the exchange-correlation function, incorporating the Perdew-Burke-Ernzerhof exchange-correlation functional [30]. A kinetic energy cutoff of 500 eV was set for the plane-wave basis set. The Brillouin zone (BZ) was sampled using the Monkhorst-Pack method [31] with a Gamma-centered k-point grid of 9 × 9 × 6. The experimental crystallographic data were adopted to perform static calculations on $Bi_2Ta_3S_6$ both without and with SOC.

## Results and Discussion

$Bi_2Ta_3S_6$ crystals have size up to 6 mm × 4 mm × 0.05 mm with metallic luster in shape of hexagonal flake, as shown in the inset of Fig. 1(a). The PXRD pattern of $Bi_2Ta_3S_6$ single crystal, as illustrated in Fig. 1(a), only shows the (00*l*) (*l*, even number) diffraction peaks, indicating the natural cleavage plane as the *ab* plane. The refinement results of SCXRD data reveal that $Bi_2Ta_3S_6$ crystallizes in a *P*6$_3$/*mcm* (No. 193) space group, with lattice parameters $a = b$ = 5.7206(3) Å, $c$ = 17.5014(13) Å, $\alpha = \beta$ = 90 °, and $\gamma$ = 120 °, aligning with previous report on powder sample [17]. Detailed crystallographic data are illustrated in Table SI [32]. Fig. 1(b) and (c) depicts the side view and top view of the crystal structure, respectively. Bi ions are intercalated in between Ta-S layers and located at the top of S site. The distances between atoms in $Bi_2Ta_3S_6$ are illustrated in Table SII [32]. Bi atoms occupy $4d$ Wyckoff positions, leaving vacancies in $2b$ sites compared to $PbTaS_2$ ($P6_3/mmc$), where Pb atoms take $2a$ Wyckoff positions [23]. As shown in Fig. 1(d), each Ta atom coordinates with six S atoms, forming a trigonal prisma where S locates in each vertex and Ta takes the geometric center. The bond angle of S1-Ta-S2 is 83.1(3) °. The adjacent Ta-S layers are arranged in a sequence analogous to that in $2H$-$TaS_2$, but exhibits a shift of 1/3 $a$ along the $a$ axis. S atoms in adjacent Ta-S layers are aligned along the $c$ axis as the formed Bi-S bond forces S atoms into $c$-axis-aligned positions, distinct from $2H$-$TaS_2$, where Ta atoms are aligned along the $c$ axis. The atomic coordinates of $Bi_2Ta_3S_6$ are shown in Table SIII [32].

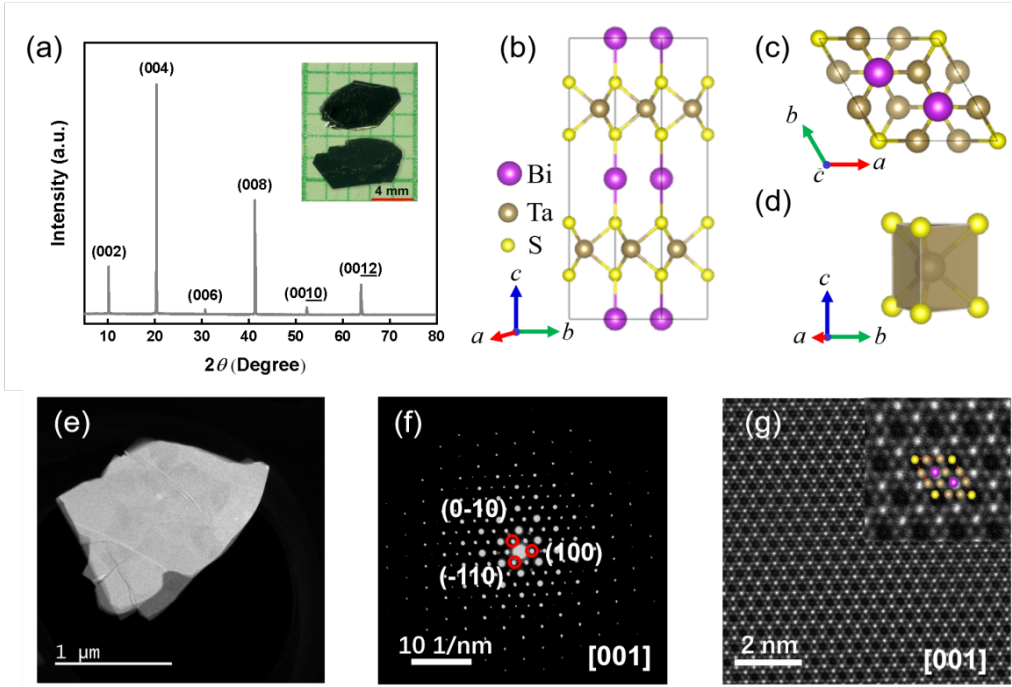

FIG. 1. The crystal structure of $Bi_2Ta_3S_6$. (a) The PXRD pattern illustrates only the (00$l$) ($l$, even number) diffraction peaks. The inset shows an optical image of as-grown crystals. (b) The side view and (c) the top view of the crystal structure of $Bi_2Ta_3S_6$, with purple spheres for Bi, dark yellow spheres for Ta, and light yellow spheres for S. (d) The schematic diagram of the trigonal prismatic configuration of Ta and S atoms. (e) The low-magnification TEM image of as-prepared specimen. (f) The SAED pattern along the $c$ axis. (g) The HAADF-STEM image of exfoliated $Bi_2Ta_3S_6$ nanosheet.

To further verify the crystal structure of $Bi_2Ta_3S_6$, we performed TEM characterizations on $Bi_2Ta_3S_6$. The low-magnification TEM image was presented in Fig. 1(e). Fig. 1(f) shows the selected area electron diffraction (SAED) pattern along the zone axis of [001] with the indexed diffraction spots of (0 -1 0), (-1 1 0), and (1 0 0) faces. The aberration-corrected HAADF-STEM image viewed along the $c$ axis illustrates the hexagonal Bi atomic layers and $TaS_2$ layers in alignment with refined crystal structure, as plotted in Fig. 1(g), which reassures the accuracy of our refinement results. SEM image in Fig. S1(a) [32] shows the morphology of $Bi_2Ta_3S_6$ single crystal. EDS shows a chemical composition Bi : Ta : S = 0.65 : 1.00 : 1.98, confirming the stoichiometric consistency with $Bi_2Ta_3S_6$. The elemental mappings indicate the homogeneous distribution of Bi, Ta, and S.

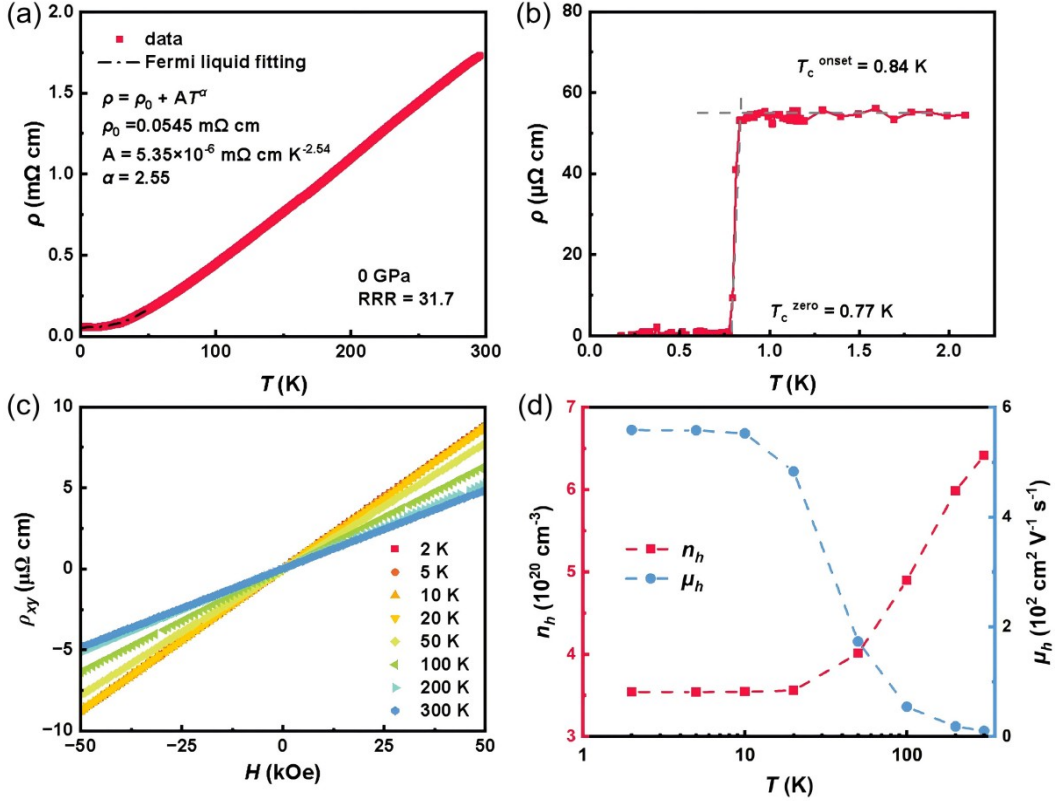

FIG. 2. The electrical transport of $Bi_2Ta_3S_6$ single crystal under ambient pressure. (a) The $\rho(T)$ curve measured from 2 K to 300 K, with the power-law fitting below 50 K. (b) The $\rho(T)$ curve in a temperature range of 0.3 K - 2 K, with the superconducting onset temperature defined as the intersection of the linear extensions of $\rho(T)$ curve in normal state and slope of superconducting transition. (c) The Hall resistivity as a function of magnetic field at temperature ranging from 2 K to 300 K. (d) The temperature evolution of $n_h$ and $\mu_h$ derived from the Hall resistivity.

Fig. 2(a) displays the in-plane temperature-dependent resistivity $\rho(T)$, which exhibits a typical metallic behavior. The low-temperature $\rho(T)$ curve can be well fitted by the power-law fitting $\rho = \rho_0 + AT^\alpha$, yielding a residual resistivity $\rho_0$ as 54.5 μΩ cm, $A$ as $5.35 \times 10^{-6}$ mΩ cm $K^{-2.55}$, and $\alpha$ as 2.55. The power $\alpha$ is close to 2, suggesting a Fermi-liquid behavior dominated by electron-phonon scattering at low temperature [33]. The determined residual resistivity ratio ($RRR = \rho(300\ K)/\rho(1.5\ K)$) is 31.7, which indicates its good crystal quality compared to the related compounds, such as $In_{0.5}TaS_2$ ($RRR = 11.5$) [15] and $TaSnS_2$ ($RRR = 10.16$) [16]. Fig. 2(b) depicts the $\rho(T)$ curve below 2 K, which shows a sharp drop to zero, indicating a superconducting transition. The superconducting onset temperature $T_c^{onset} = 0.84$ K, and the zero resistivity is achieved at $T_c^{zero} = 0.77$ K. As plotted in Fig. 2(c), the Hall resistivity, $\rho_{xy}$, exhibits a linear field dependence across all temperatures, which implies a single band nature. The positive slop indicates that the dominant carrier type is hole. At 2 K, the Hall coefficient $R_H$ is about $1.8 \times 10^{-10}$ m$^3$/C. Based on the Drude model, the corresponding carrier concentration $n_h$ is determined to be $3.54 \times 10^{20}$ cm$^{-3}$ with the carrier mobility $\mu_h$ as 588 cm$^2$ V$^{-1}$ s$^{-1}$. While at 300 K, the determined carrier mobility is about 10 cm$^2$ V$^{-1}$ s$^{-1}$. The obtained carrier concentration of Bi$_2$Ta$_3$S$_6$ is rather low compared to other ATaX$_2$ compounds such as PbTaSe$_2$ [20], PbTaS$_2$ [23], and In$_{0.58}$TaS$_2$ [14], showing the limited contribution from Bi to carriers. Fig. 2(d) illustrates the temperature-dependent $n_h$ and $\mu_h$ extracted from the Hall resistivity. The monotonically increasing $n_h$ alongside decreasing $\mu_h$ as temperature rises from 2 K to 300 K, aligns with typical metallic transport behavior.

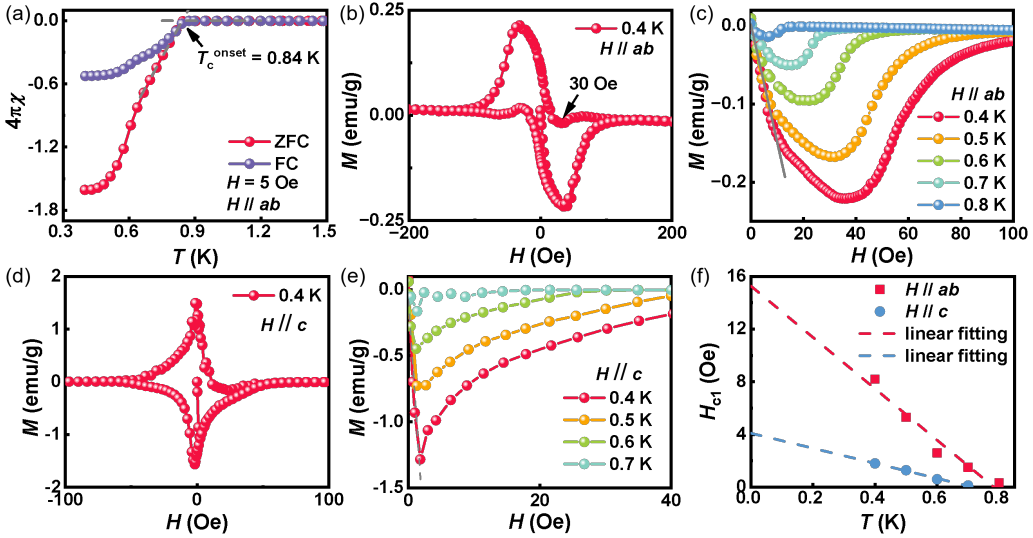

FIG. 3. The anisotropic magnetic properties of Bi$_2$Ta$_3$S$_6$ single crystal under ambient pressure. (a) The $\chi(T)$ curves measured under a magnetic field of 5 Oe with ZFC (red dots) and FC (blue dots) modes. $T_c^{onset}$ is identified as the intersection of the extensions of normal-state $\chi(T)$ and the slope of diamagnetic signal. The *M-H* curves at 0.4 K with (b) *H//ab* and (d) *H//c*, showing magnetization hysteresis loops. The *M-H* curves at temperature ranging from 0.4 K to 0.8 K with (c) *H//ab* and (e) *H//c*. (f) The linear fittings of anisotropic $H_{c1}$ at various temperature.

Fig. 3(a) demonstrates the temperature dependence of zero field cooling (ZFC) and field cooling (FC) magnetic susceptibility ($\chi$) under a magnetic field (*H*) of 5 Oe, which reveals $T_c$ = 0.84 K in agreement with the aforementioned result. The superconducting volume fraction estimated from ZFC curve at 0.4 K is larger than 100%, indicating a bulk superconductivity. Fig. 3(b) and (d) shows the field-dependent magnetization curves at 0.4 K with the in-plane magnetic field (*H//ab*) and out-of-plane magnetic field (*H//c*), respectively. The Messiner state and mixed state can be well distinguished in both *M-H* curves, suggesting its type-II superconducting nature. For *H//ab*, a slight downward concavity around 30 Oe signals vortex pinning, as denoted by a black arrow in Fig. 3(b). Fig. 3(c) and (e) shows the *M-H* curves at varying temperature with *H//ab* and *H//c*, respectively. The lower critical magnetic field $H_{c1}$ is determined by the intersection between the extension of linear *M-H* curve in Meissner state and the *x*-axis, as plotted in Fig. 3(f). Due to the limited data on ultra-low temperature measurements, a liner fitting is adopted to estimate anisotropic $H_{c1}(0)$, which is 15 Oe for *H//ab* and 4 Oe for *H//c*, respectively.

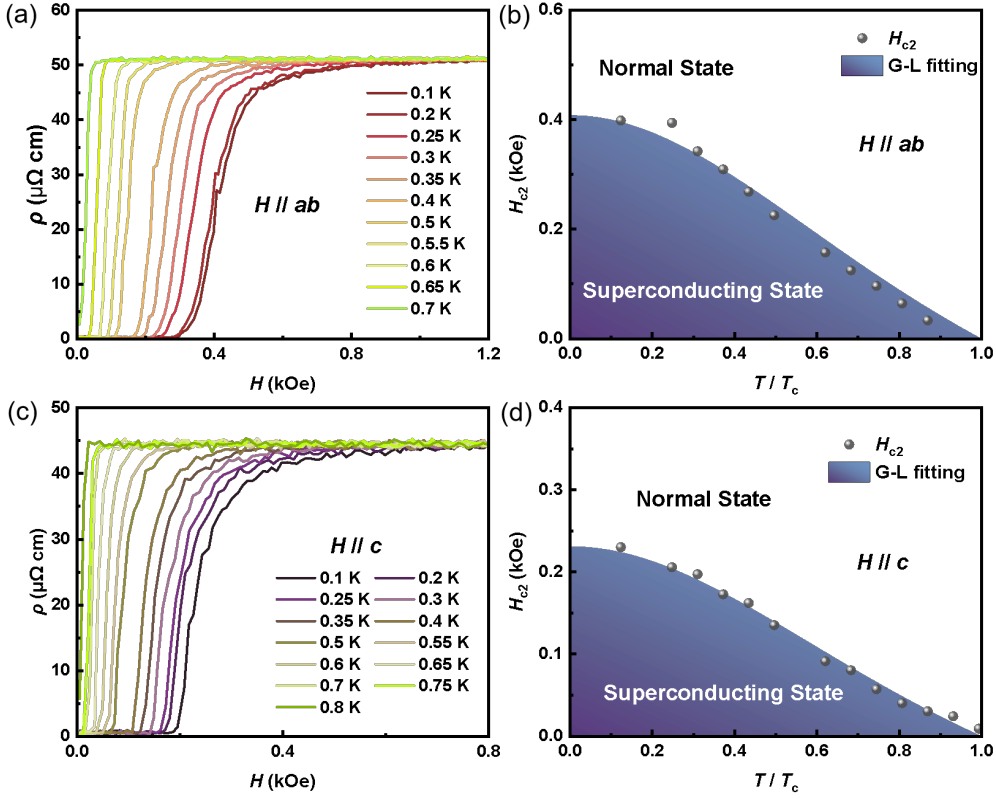

FIG. 4. The anisotropic $H_{c2}$-*T* phase diagram of $Bi_2Ta_3S_6$. The field-dependent resistivity at temperature ranging from 0.1 K to 0.7 K with *H//ab* (a) and from 0.1 K to 0.8 K with *H//c* (c). The temperature-dependent $H_{c2}$ with *H//ab* (b) and *H//c* (d). The gray dots represent $H_{c2}$ at varying temperatures.

Fig. 4(a) and (c) present the field-dependent resistivity at different temperatures with *H//ab* and *H//c*, respectively. The upper critical magnetic field $H_{c2}$ was defined as the magnetic field at which the resistivity reaches 50% of the normal-state value. With increasing temperature, both upper critical fields with *H//ab* ($H_{c2}^{ab}$) and *H//c* ($H_{c2}^{c}$) exhibit a monotonic decrease. The *H-T* phase diagrams where $H_{c2}$ is plotted as a function of normalized temperature ($T/T_c(0)$) with *H//ab* and *H//c*, are illustrated in Fig. 4(b) and (d), respectively. $T_c(0)$ was taken at a temperature of 50% resistivity drop at zero magnetic field as 0.805 K. The Ginzburg-Landau (G-L) fittings were adopted, yielding the zero-temperature upper critical field with *H//ab* ($H_{c2}^{ab}(0)$) as 404 Oe and that with *H//c* ($H_{c2}^{c}(0)$) as 231 Oe. Both $H_{c2}(0)$ values lie below the BCS Pauli paramagnetic limit ~ 1.83 $T_c$ as 1.54 T, classifying $Bi_2Ta_3S_6$ as a weakly coupled superconductor. The upper critical field anisotropy parameter is calculated as $\gamma = H_{c2}^{ab}/H_{c2}^{c} = 1.74$, which is smaller than that of $TaSnS_2$ [16], $PbTaSe_2$ [22], and $Pb_{1/3}TaS_2$ [34]. The anisotropic G-L parameter $\kappa$ with *H//ab* and *H//c* is determined as $\kappa_{ab}$ = 7.67 and $\kappa_c$ = 4.50, much larger than $1/\sqrt{2}$, which further testify $Bi_2Ta_3S_6$ as a type-II superconductor. The anisotropic superconducting coherence length $\xi_{ab}$ is 37.7 nm, and $\xi_c$ is 21.6 nm. Furthermore, the anisotropic penetration depth $\lambda$ can be determined as $\lambda_{ab}$ = 289.2 nm and $\lambda_c$ = 57.0 nm. $Bi_2Ta_3S_6$ shows the lowest $T_c$ and $H_{c1}$ among all known $ATaS_2$-type superconductors, with superconducting parameters summarized in Table SIV [32].

Hydrostatic pressure was applied to study the evolution of superconductivity with pressure, where $Bi_2Ta_3S_6$ flake was placed in a horizontal position. Fig. 5(a) shows the $\rho(T)$ curves under pressure up to 13.5 GPa. As pressure increases, the resistivity at 300 K gradually declines. The anomaly around 240 K at 3 GPa may caused by a phase transition of PTM. The inset shows the enlargement of $\rho(T)$ curves below 6 K, where obvious superconducting transitions can be distinguished. In the beginning, both $T_c^{onset}$ and $T_c^{zero}$ are dramatically enhanced under hydrostatic pressure. $T_c^{onset}$ reaches its maximum of 3.42 K at 5 GPa, which is 4 times as that at ambient pressure, and is relatively stable above 5 GPa. At 13.5 GPa, $T_c^{onset}$ is 3.06 K, slightly lower than that at 5 GPa. No sign of total suppression of the superconductivity is observed within the studied pressure range. The evolution of $T_c^{onset}$ and $T_c^{zero}$ with pressure are summarized in Fig. 5(c). To further inspect the evolution of zero-temperature upper critical field $H_{c2}(0)$ with pressure, the temperature-dependent resistivity was measured under varying magnetic field and pressure with $H//c$, as depicted in Fig. S2 [32]. Under each pressure, as magnetic field increases, $T_c$ gradually decreases due to the pair-breaking effect. Taking temperature of 50% value of the normal-state resistivity during the transitions as $T_c(H)$, $H_{c2}(0)$ was extracted via G-L fittings, as plotted in Fig. 5(b). As illustrated in Fig. 5(c), $H_{c2}(0)$ shows enhancement from 0 GPa to 5 GPa and reaches the maximum of 1.06 T at 5 GPa,, which is about 4.5 times larger than that at ambient pressure. The concurrent maxima in $H_{c2}(0)$ and $T_c$ coincides with the abrupt change in preceding fitting parameters at 5 GPa, which may be attributed to a potential pressure-induced phase transition worth further investigation. Subsequently, $H_{c2}(0)$ undergoes a gradual suppression beyond 5 GPa, and reaches 0.51 T at 12 GPa.

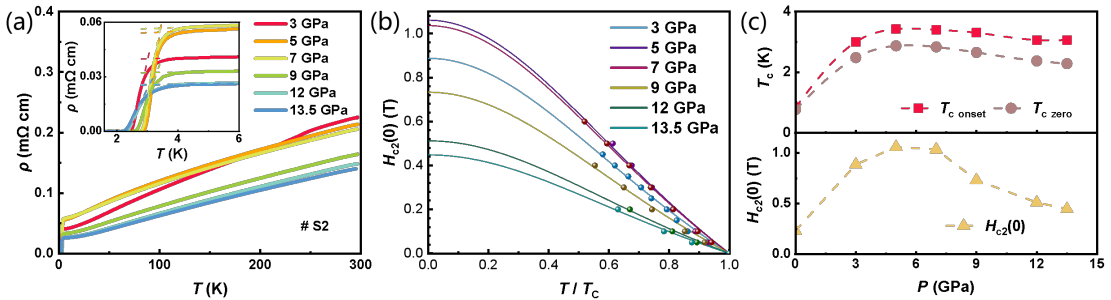

FIG. 5. The electrical transport properties of $Bi_2Ta_3S_6$ single crystal under pressure. (a) The $\rho(T)$ curves at temperature ranging from 1.8 K to 300 K under pressure from 3 GPa to 13.5 GPa. The inset shows the enlargement of $\rho(T)$ curves below 6 K. (b) The G-L fittings of temperature-dependent upper critical field $H_{c2}$ under various pressure. (c) The evolution of superconducting transition temperature and zero-temperature upper critical field with pressure.

As depicted in Fig. 6(a), the $\rho(T)$ curves in normal state below 50 K under pressure were analyzed by power-law fitting. The obtained parameters as a function of pressure are summarized in Fig. 6(b). As pressure increases, $\rho_0$ descends below 3 GPa and ascends from 3 GPa to 7 GPa. At 7 GPa, $\rho_0$ reaches 0.056 mΩ cm, and decreases again above 7 GPa. The coefficient $A$ increases below 5 GPa, and then decreases from 5 GPa to 13.5 GPa. The power $\alpha$ experiences a drop from 2.54 at ambient pressure to 1.16 at 5 GPa, following a steady rise to 1.71 up to 13.5 GPa.

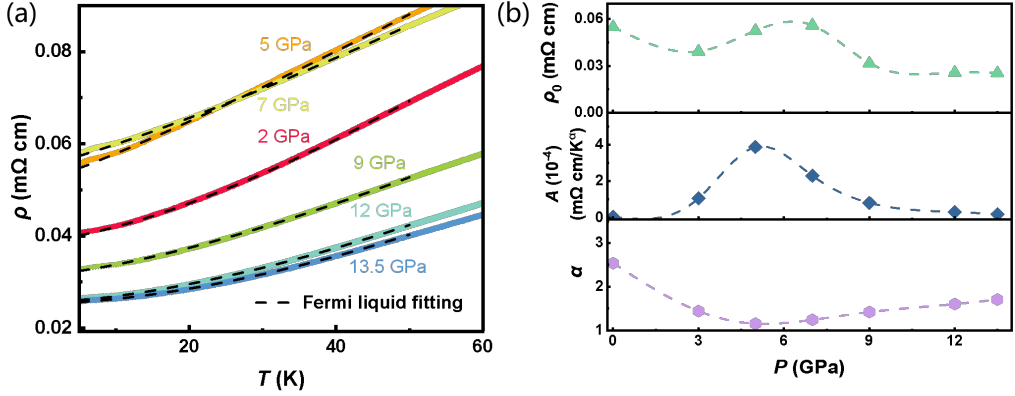

FIG. 6. The electrical transport properties in the normal state of $Bi_2Ta_3S_6$ below 60 K under pressure. (a) The $\rho(T)$ curves under pressure at temperature ranging from 5 K to 60 K. The dashed lines represent the power-law fittings of $\rho(T)$ below 50 K. (b) The evolution of power-law fitting parameters under increasing pressure.

To examine topological nature of the electronic structure of $Bi_2Ta_3S_6$, DFT calculations at ambient pressure have been performed. Fig. 7(a) shows the bulk electronic structure without SOC, where a continuous gap except the K and H points (including K-H line) is observed. As a layered compound, the in-plane dispersions show that there is a Dirac point at K/H point. The detailed fatted bands show that the states of the Dirac point are mainly from the Bi-$p_x$, Bi-$p_y$ orbitals in the honeycomb layer. As the unit cell of $Bi_2Ta_3S_6$ has two Bi-honeycomb layers, there are actually two Dirac points at K/H point. The weak $k_z$-dispersive states of the Dirac points indicate the interlayer hopping between the adjacent Bi-honeycomb layers. Once including SOC, the Dirac points at K/H point become gapped as shown in Fig. 7(b). With the inversion symmetry, the topology of the gapped band structure is described by the ($Z_2,Z_2,Z_2; Z_4$) symmetry indicators. The computed indices (0,0,0; 2) indicate that it is topologically nontrivial.

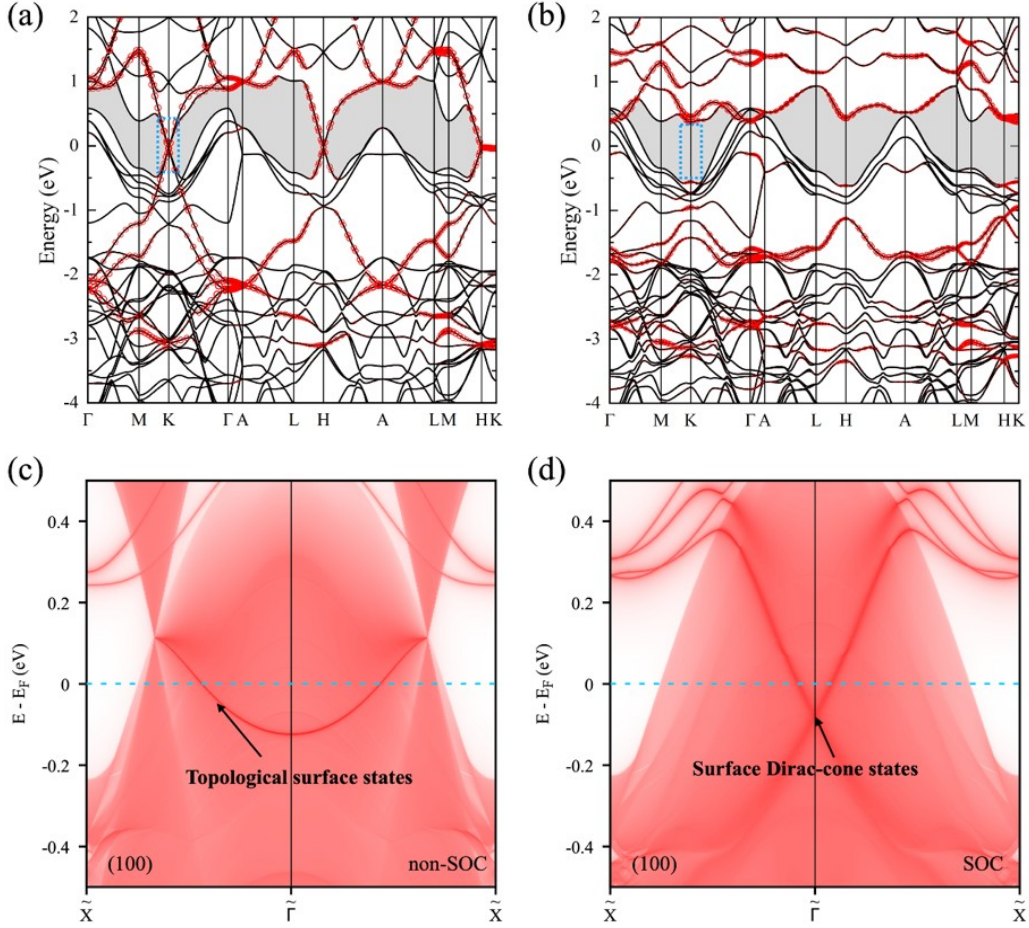

FIG. 7. DFT calculations of the electronic structure of $Bi_2Ta_3S_6$. Electronic band structure (a) without and (b) with SOC, showing a Dirac point at the K/H point (blue box). The size of the red circles indicates the weight of Bi-$p_{x,y}$ orbitals. Once including SOC, the Dirac points become gapped, and the band overlap at Γ is removable. The (100)-surface spectra (c) without and (d) with SOC. In panel (c), topological surface states are connecting the projections of the bulk Dirac points. Instead, the topological surface Dirac-cone states are clearly shown in panel (d).

As a three-dimensional stacking of quantum spin Hall insulator layers, the topological Dirac states are expected on the (100) surface. For this purpose, we constructed the tight-binding Hamiltonian with Ta-$d$, Bi-$p$, and S-$p$ orbitals via the Wannier90 software package [35]. We used an iterative method to obtain the surface Green's function of the semi-infinite system, as employed in the WannierTools [36]. The imaginary part of the Green's function is the local density of states at the surface. The surface states on the (100) surface without and with SOC are shown in Fig. 7(c) and 7(d). In the absence of SOC, the topological surface states are clearly shown, which are connecting the projections of the bulk Dirac points. In the presence of SOC, the surface Dirac-cone states are clearly shown at $\bar{\Gamma}$. Although these surface states are buried within the projections of the bulk states, they remain weakly coupled, as the topological surface states originate from Bi-$p_{x,y}$ orbitals (Bi-honeycomb layer), while the bulk states are from the Ta-$d$ orbitals ($TaS_2$ layer). Conclusively, $Bi_2Ta_3S_6$ having a superconducting transition at 0.84 K shows nontrivial topological surface states on (100) surface, which gives us more clues to discover new potential topological superconductor in ATaX$_2$-type materials.

## Conclusion

Bi$_2$Ta$_3$S$_6$ is a typical 2D material crystallizing in hexagonal space group $P6_3/mcm$ (No. 193), where Bi atoms are intercalated in between two Ta-S layers. A sharp superconducting transition is observed at 0.84 K, similar to that in 2$H$-TaS$_2$. The magnetization of Bi$_2$Ta$_3$S$_6$ confirms its nature as a type-II superconductor. The $H_{c2}(0)$ under out-of-plane magnetic field at ambient pressure is 404 Oe, far smaller than the Pauli limit ~ 1.84 $T_c$, which suggests Bi$_2$Ta$_3$S$_6$ is a weakly-coupled superconductor. The DFT calculations reveal that the bands near the Fermi level are primarily contributed by the $p$-orbit of Bi atoms and $d$-orbit of Ta atoms, and the surface states on the (100) surface shows two Dirac points. The intercalation of Bi between S-Ta-S layers seems causing limited effect on the carrier density compared to other ATaX$_2$-type materials, yet leads to nontrivial topological surface states. The intrinsic type-II superconductivity coupled with predicted topological surface states positions Bi$_2$Ta$_3$S$_6$ as a candidate for topological superconductors, which requires subsequent spectroscopic verification.

## Acknowledgements

C. L. Li and G. Wang would like to thank Prof. X. L. Chen of the Institute of Physics, Chinese Academy of Sciences and Dr. Z. N. Guo of the University of Science and Technology Beijing for fruitful discussions. This work was supported by the National Key Research and Development Program of China (Grant No. 2022YFA1403900, Grant No. 2022YFA1403800, and Grant No. 2023YFA1607400), the National Natural Science Foundation of China (Grant No. 52325201 and Grant No. 12188101), and the Center for Materials Genome. This work was also partially supported by the ultra-low temperature high pressure physical property measurements - cubic Anvil cell station (A2) and sample pre-selection and characterization station (F2) of Synergetic Extreme Condition User Facility, Beijing, China.